\documentclass[final,12pt,draft]{elsarticle}
\usepackage{graphicx}
\usepackage{amsmath} 
\usepackage{amssymb}
\biboptions{numbers,sort&compress}
\usepackage{siunitx}
\usepackage{color} 
\usepackage{booktabs,caption}

\def\figref#1{Fig.~\ref{#1}}
\def\tabref#1{Tab.~\ref{#1}}
\def\eqnref#1{Eqn.~(\ref{#1})}
\journal{Journal of Luminescence} 
\begin{document}

\begin{frontmatter}
\title{{\bf Cathodo- and radioluminescence of 
 Tm$^{3+}$:YAG and Nd$^{3+}$:YAG in an extended wavelength range}}
\author[mymainaddress]{A. F. Borghesani\corref{mycorrespondingauthor}}
\address[mymainaddress]{CNISM Unit, Department of Physics and Astronomy, University of Padua and\\
Istituto Nazionale Fisica Nucleare, sez. Padova\\ via F-Marzolo 8, I-35131 Padua, Italy}\cortext[mycorrespondingauthor]{Corresponding author}
\ead{armandofrancesco.borghesani@unipd.it}

\author[mytertiaryaddress]{C. Braggio}

\author[mysecondaryaddress]{G. Carugno}
\address[mysecondaryaddress]{Istituto Nazionale Fisica Nucleare, sez. Padova and \\ Department of Physics and Astronomy, University of Padua\\via F-Marzolo 8, I-35131 Padua, Italy}
\author[mytertiaryaddress]{F. Chiossi}
\address[mytertiaryaddress]{Department of Physics and Astronomy, University of Padua and \\
Istituto Nazionale Fisica Nucleare, sez. Padova \\via F-Marzolo 8, I-35131 Padua, Italy}

\author[mytertiaryaddress]{M. Guarise}

\begin{abstract}
We have studied the cathodo- and radioluminescence of Nd:YAG and of Tm:YAG single crystals in an extended wavelength range up to $\approx 5\,\mu$m in view of developing a new kind of detector for low-energy, low-rate energy deposition events. Whereas the light yield in the visible range is as large as $\approx 10^{4}\,$photons/MeV, in good agreement with literature results, in the infrared range we have found a light yield $\approx 5\times 10^{4}\,$photons/MeV, thereby proving that ionizing radiation is particularly efficient in populating the low lying levels of rare earth doped crystals.
\end{abstract}


\end{frontmatter}

\section{Introduction}\label{sec:intro}

In our laboratory we are developing a new kind of scintillation detectors to investigate low-rate, low-energy deposition events. We have decided to adopt the so called 
InfraRed Quantum Counter (IRQC) scheme, initially proposed by Bloembergen as early as 60 years ago~\cite{Basov1959}. In the IRQC scheme the intrinsic limitations of traditional infrared detection are overcome by shifting the detection in the visible range.

This scheme requires that the active material of the detector has, at least, three energy levels: the ground state, an intermediate low-energy level, and a high-energy one. The particle to be detected excites the material from the ground state into the intermediate level. The population of this level is promoted to the highly-lying level by means of a suitably tuned laser. Finally, the high-lying state radiatively relaxes to the ground state. 
visible fluorescence is then easily detected with usual techniques. Promising candidates for the active material of the detector are Rare Earth (RE) doped crystals because the upconversion processes are highly efficient and are also well studied~\cite{auzel2004}. The possibility to apply the IRQC scheme for particle detection has been demonstrated in Er-doped YAG single crystal~\cite{borghesani2015}.

A key assumption for the development of an upconversion-based detector is that the particle energy loss in the material originates a wideband excitation that more efficiently contributes to the population of the low-lying energy levels rather than the higher-lying ones. We then expect that 
the cathodo- and radioluminescence spectra display a large infrared component. In this way, this new kind of detector should be endowed of an improved efficiency and energy resolution with respect to the most commonly used solid-state inorganic scintillation detectors.

A quantitative validation of this hypothesis cannot be obtained by literature. Actually, the infrared luminescence is hardly investigated because of the long lifetime of the low-lying levels involved~\cite{moses1998,antonini2002}. In fact, quantitative studies have been carried out in the visible range by exciting RE-doped crystals with different ionizing radiations, e.g., $\gamma-$ and $X-$rays, and neutrons (for a review, see~\cite{yanagida2013}), ion beams~\cite{brooks2002,khanlary2005}, $\alpha -$ particles~\cite{yanagida2011}, and synchrotron radiation~\cite{Ning2007}, whereas only optical spectroscopic investigations are available in the infrared range~\cite{eichhorn2008}.

In order to check the validity of our assumption, we have started a systematic study of the luminescence properties of RE-doped crystals excited by energetic electrons and X-rays, focusing on the infrared component of the emission. 
The quantitative analysis of the infrared component we report in this work is based on the comparison with the visible counterpart. The good agreement 
of our results in the visible range with literature data lends credibility to the quantitative results we get in the infrared range.

We show here the first results of this comprehensive study on cathodo- (CL) and radioluminescence (RL) in Nd:YAG and in Tm:YAG single crystal samples from deep UV (DUV) up to mid IR ($\approx 5\,\mu$m).

\section{Experimental Method and Apparatus}\label{sect:app}
The experimental method consists in exciting the crystal under investigation by either high-energy electrons or X-rays. 
The induced CL and RL are then analyzed in a very wide wavelength band covering the range from DUV $\lambda \approx 200\,$nm up to mid infrared $\lambda\approx 5\,\mu$m. 

\begin{figure}[t!]
    \centering
    \includegraphics[width=\textwidth]{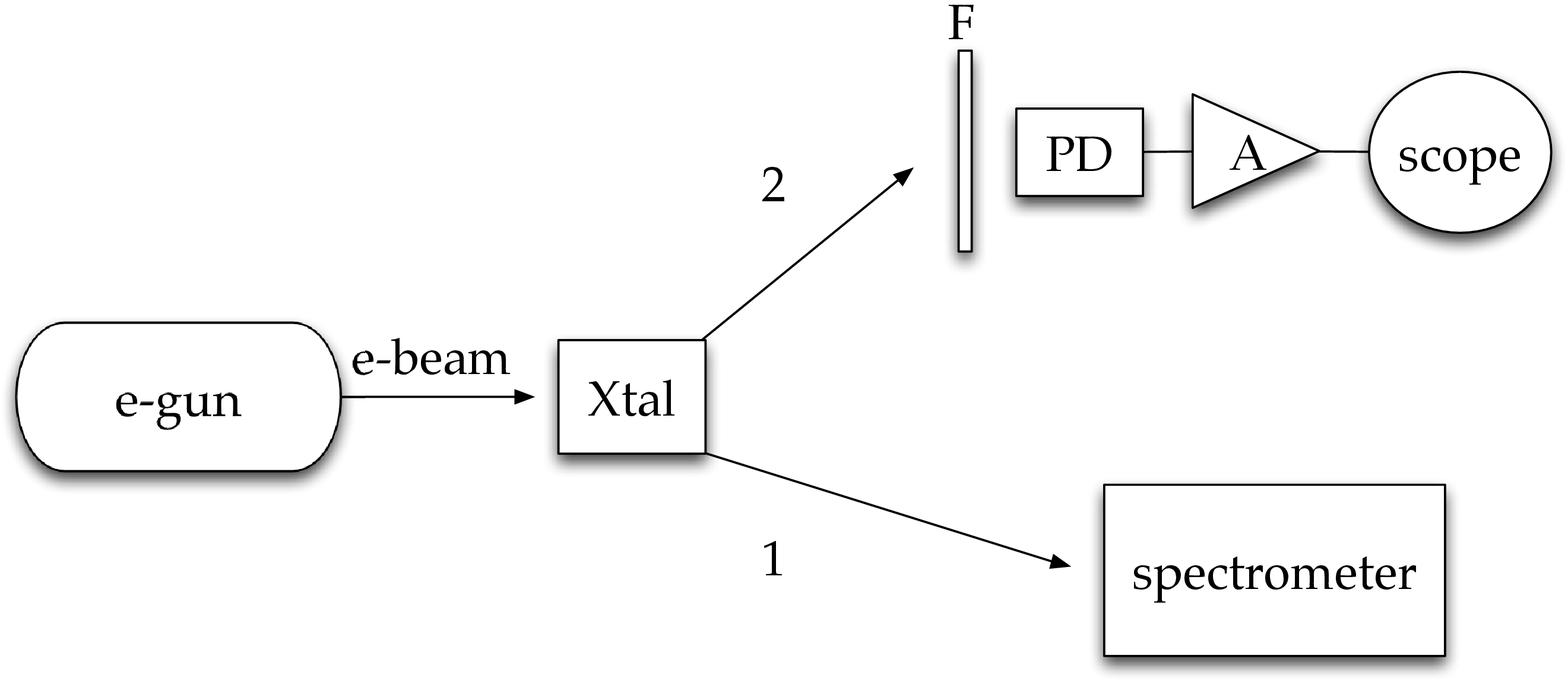}
    \caption{\small Simplified scheme of the experimental apparatus. Xtal: crystal. F: Filter. PD: Photodiode. A:Amplifier. e-gun: electron gun. e-beam:electron beam. \label{fig:app}}
\end{figure}

The apparatus built for implementing this method is schematically shown in \figref{fig:app}.
Electrons of energy up to $100\,$keV are produced by a home-made electron gun (e-gun) that has been thoroughly described elsewhere~\cite{barcellan2011}.  We briefly recall here only its main features. 
It can be operated in continuous or pulsed mode. In the continuous mode, it can deliver current of intensity up to $15\,\mu$A on the crystal. 
 In pulsed mode, electrons are supplied in bunches containing up to a few tens of nC with a pulse duration adjustable at will between a few tens of $\mu$s up to the ms-range. It is also possible to vary the pulse repetition rate in the range $20-1000\,$Hz when the radiative lifetime of the crystal levels is measured. In both modes, the electron energy is set at $70\,$keV.
\begin{figure}[b!]
\centering
\includegraphics[width=0.5\textwidth]{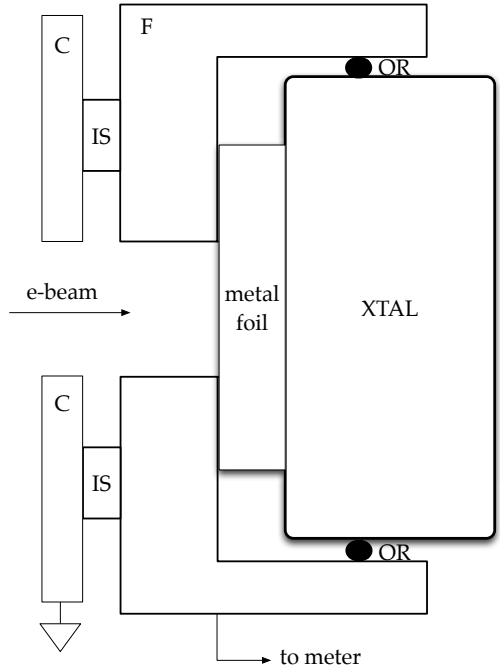}
\caption{\small Detail of the crystal mounting (not to scale). XTAL= crystal. Metal foil: $10\,\mu$m-thick Ti (or Ta for X-ray production) foil. F: Flange. IS: Insulating spacers. C: grounded collimator plate. OR: O-ring.
\label{fig:montaggioXtal}}
\end{figure}

The electron beam is collimated to a spot area of $\approx 3\,$mm$^{2}$. The crystal  is mounted on an insulated flange, as shown in \figref{fig:montaggioXtal}, and is in electrical contact with it by means of a $\approx 10\,\mu$m thin metal foil. Crystal and flange act as a Faraday cup (or, beam stopper) that allows us to measure the amount of injected charge. Ti is used for the metal foil if CL has to be studied. Its small atomic number $Z$ allows the electrons to cross the foil  and to impinge on the crystal surface with $15 \,$keV energy loss~\cite{estar} and to be immediately recollected on the Faraday cup for the charge measurement. 
On the other hand, if the high $Z$, thin Ta foil is used, electrons are stopped in the metal foil and X-rays are produced, whose intensity is proportional to the amount of the injected charge. In this way, RL can be investigated.

We used two commercial YAG single crystals doped with Nd$^{3+}$
 ($1.1\, \%$ at.) and with Tm$^{3+}$ ($4.4\, $\% at.), 
respectively. They are shaped as small cylinders of $3\,$mm in height. Their diameters are $3\,$mm for Nd:YAG and $5\, $mm for Tm.
The luminescence produced by the crystal can be spectrally analyzed with the aid of suitable spectrometers (path 1 in \figref{fig:app}). For the $200-1000\,$nm range, 
we have used Si CCD spectrometer (OceanOptics, mod. RedTide 650). For the $900-1700\,$nm range, we used the InGaAs CCD spectrometer (OceanOptics, mod. NIR 512).  For the $2000-12000\,\mbox{cm}^{-1}$ wavenumber range, we used a Fourier-Transform Infrared interferometer (Bruker, mod. Equinox 55) equipped with either InGaAs, InAs, or InSb detectors depending on the investigated spectral ranges. We have devised a procedure, described in the Appendix, to reliably merge spectra obtained with spectrometers spanning different spectral ranges.

Our apparatus is also designed  to make absolute measurements of the light yield (LY), i.e., the total amount of emitted light as a function of the energy deposited into the crystal by either electrons or X-rays. To this purpose (path 2 in \figref{fig:app}), the luminescence light is collected by a photodetector PD. The PD signal is amplified by amplifier A and recorded on a storage oscilloscope. A PC fetches the data from the oscilloscope for offline analysis. The LY measurements can be restricted to selected wavelength bands by using suitable combinations of optical filters F. We used two Si- (Thorlabs, mod. Det36A and Hamamatsu, mod. S1337-1010BQ), InGaAs (Thorlabs, mod. Det20C), and a liquid nitrogen cooled InAs (Hamamatsu, mod. P7163) PDs.

The PD is mounted on a $z-$translational stage in order to change the detector-to-crystal distance and, thus, to measure the solid angle subtended by the crystal at the detector for normalization purposes.  

The output signal of the PD can be either integrated, if one is only interested to the total amount of emitted light, or can be linearly amplified if the lifetime of the radiative levels of the crystal has to be determined from the time evolution of the signal. For the former goal, we used an active integrator with time constant $\tau_{c}\approx 480\,\mu$s and conversion factor $G=0.25\,$mV/fC that makes this device particularly useful for the measurements with X-rays that produce a very faint luminescence signal. 

For the lifetime measurements, we used a transimpedance amplifier (Fem\-to, mod. DLPCA-200) with gain $G=10^{6}\,$V/A. In this condition, the amplifier bandwidth is $\approx 500\,$kHz and relatively fast PD responses are quite faithfully reproduced.  
\begin{figure}[b!]
\centering
\includegraphics[width=\columnwidth]{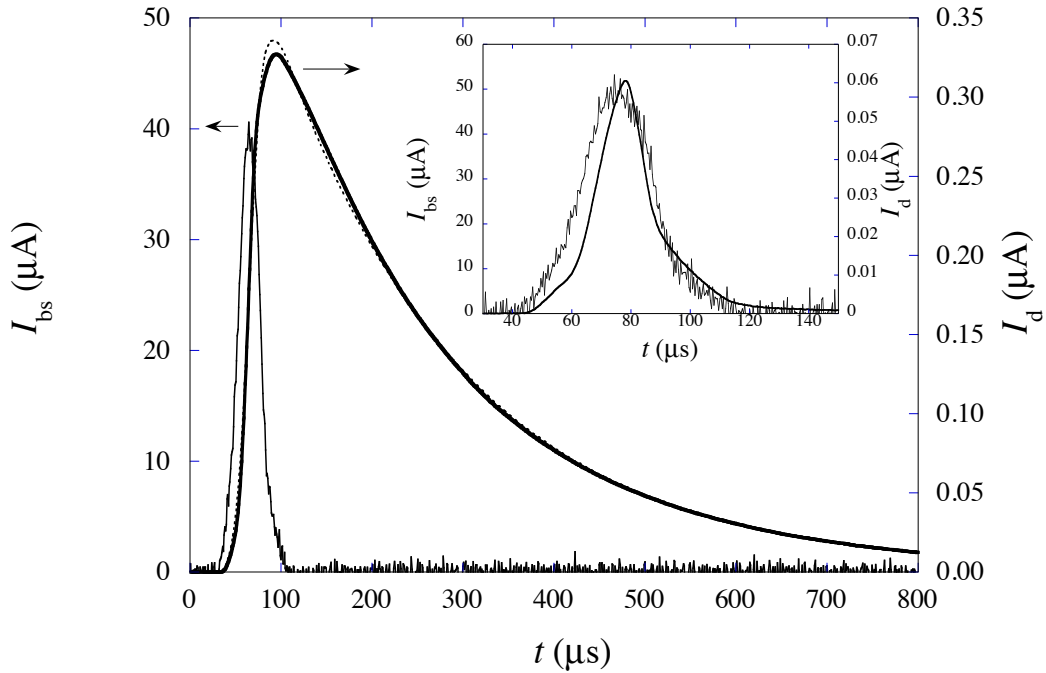}
\caption{\small Time evolution of the IR- (main plot) and visible CL (inset) of Nd:YAG (thick solid line, right scale). Thin solid line: current injection waveform (left scale). Dotted line: numerical simulation.
\label{fig:segnali}}
\end{figure}
In order to show the accuracy with which we can measure relatively long radiative lifetimes,  we report in \figref{fig:segnali} the time evolution of the infrared emission of Nd:YAG crystal excited by an electron pulse. In \figref{fig:segnali} $I_\mathrm{bs}$ is the $\approx 40\,\mu$s-long current pulse (left scale) and $I_{d}$ (right scale) is the amplified Si PD response. The dashed line is the PD response numerically computed by using the experimentally measured current injection shape, from which a value $\tau =(212\pm 10)\,\mu$s is obtained, in agreement with literature data~\cite{Venikouas1984}.

The inset in the figure shows the  PD response to the visible light originating from the decay of the  short-lived $^{2}\mathrm{F}(2)_{5/2}$ manifold  ($\tau\approx 3\,\mu$s)~\cite{Venikouas1984} excited by electron pulse.  In this case, the amplified PD response quite rapidly follows the current injection. Although the lifetime can be determined by an experimental fit of the signal long-time tail, our approach to numerically simulate the signal allow us, if necessary, to investigate the level population kinetics.

The integration of the beam stopper and PD signals yields 
the amount of charge $Q_d$ generated in the active material of the detector and the amount of charge per pulse $Q_\mathrm{bs}$ accelerated towards the crystals and collected by the Faraday cup.

\section{Experimental Results and Discussion}
The response of a detector to the luminescence light produced by a scintillator material is given by
\begin{equation}
Q_{d} =e\eta \,E_{i}\left(\frac{\Delta \Omega}{4\pi}\right)
\mathrm{LY}
\label{eq:QD}
\end{equation}
in which $e$ is the elementary charge, $\eta$
 is the detector quantum efficiency, and $E_{i}$ is the amount of energy released in the crystal by the ionizing radiation. $\Delta \Omega=S/d^{2}$ is the fraction of solid angle subtended by the detector of cross sectional area $S$ at the source located at distance $d.$ 
From \eqnref{eq:QD} the $\mathrm{LY}_i$, number of photons/MeV emitted in the optical range $\Delta \lambda_i$, can be expressed as
\begin{eqnarray}
\mathrm{LY}_i &=&   \frac{1}{k} \frac{4 \pi d^2}{S} \frac{Q_d}{Q_\mathrm{bs}}\nonumber \\
& & \left[\left(\int_{\Delta \lambda_i} I(\lambda) \lambda d\lambda \right) \bigg/ \left( \int \eta(\lambda) T(\lambda) I(\lambda)\lambda d\lambda \right) \right]
\label{eqn:LY5}
\end{eqnarray}
where $I$ is the scintillator emission spectrum and, thus, the term in square parenthesis takes into account the wavelength dependence of $\eta$ and the transmission $T$ of the optical filter possibly mounted in front of the detector. 
The constant $k= E_{i}/Q_\mathrm{bs}$ takes on different values for electron- or X-rays excitation.

\begin{figure}[t!]
\centering
\includegraphics[width=1\textwidth]{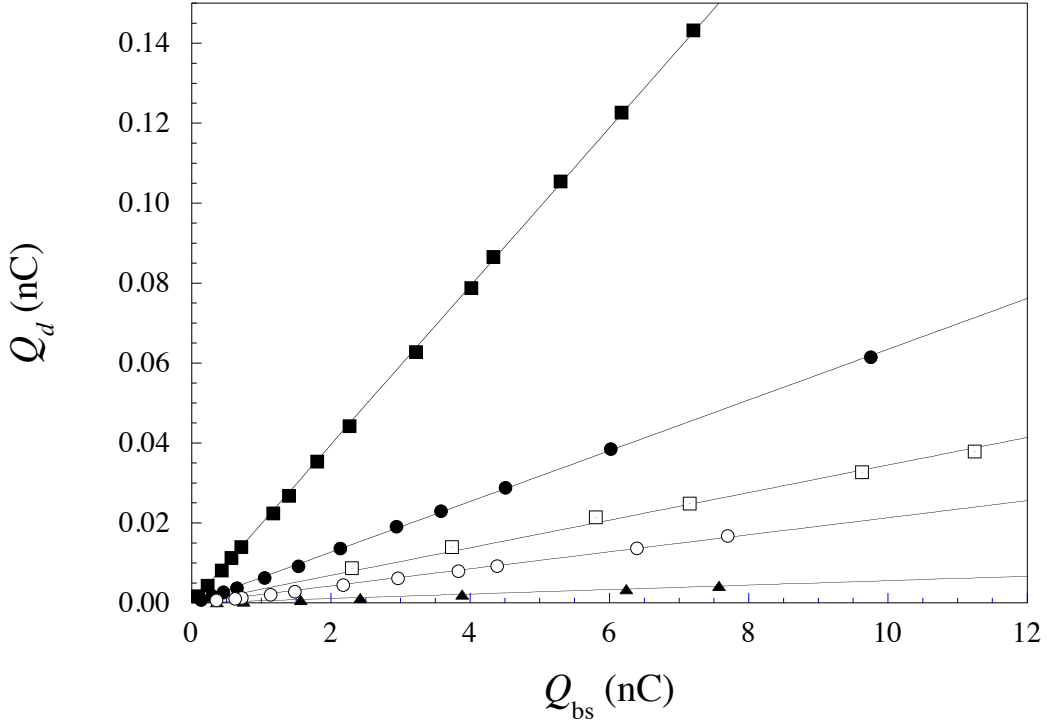}
\caption{\small 
$Q_{d}$ vs 
$Q_\mathrm{bs} $ for the electron impact excited Tm:YAG crystal in several wavelength ranges.  Closed squares:  $(200\lesssim \lambda\lesssim 1100)\,$nm. Closed circles: $\lambda =(450 \pm 40)\,$nm. Open circles: $\lambda =(850 \pm 40)\,$nm. Triangles: $\lambda =(650 \pm 40)\,$nm. Opens squares: $(1000\lesssim \lambda\lesssim 3100)\,$nm.
The error bars are of the same size of the symbols.
\label{fig:LinTmegun}}\end{figure}

In our experiment we directly measure $Q_d$ and $Q_\mathrm{bs}$ and we have verified their proportionality with all crystals in any wavelength bands, independently of the excitation source. As an example, we show in \figref{fig:LinTmegun} $Q_{d}$ as a function of $Q_\mathrm{bs}$ for the electron-beam excited Tm:YAG crystal.

\begin{figure}[t!]
\centering
\includegraphics[width=1\textwidth]{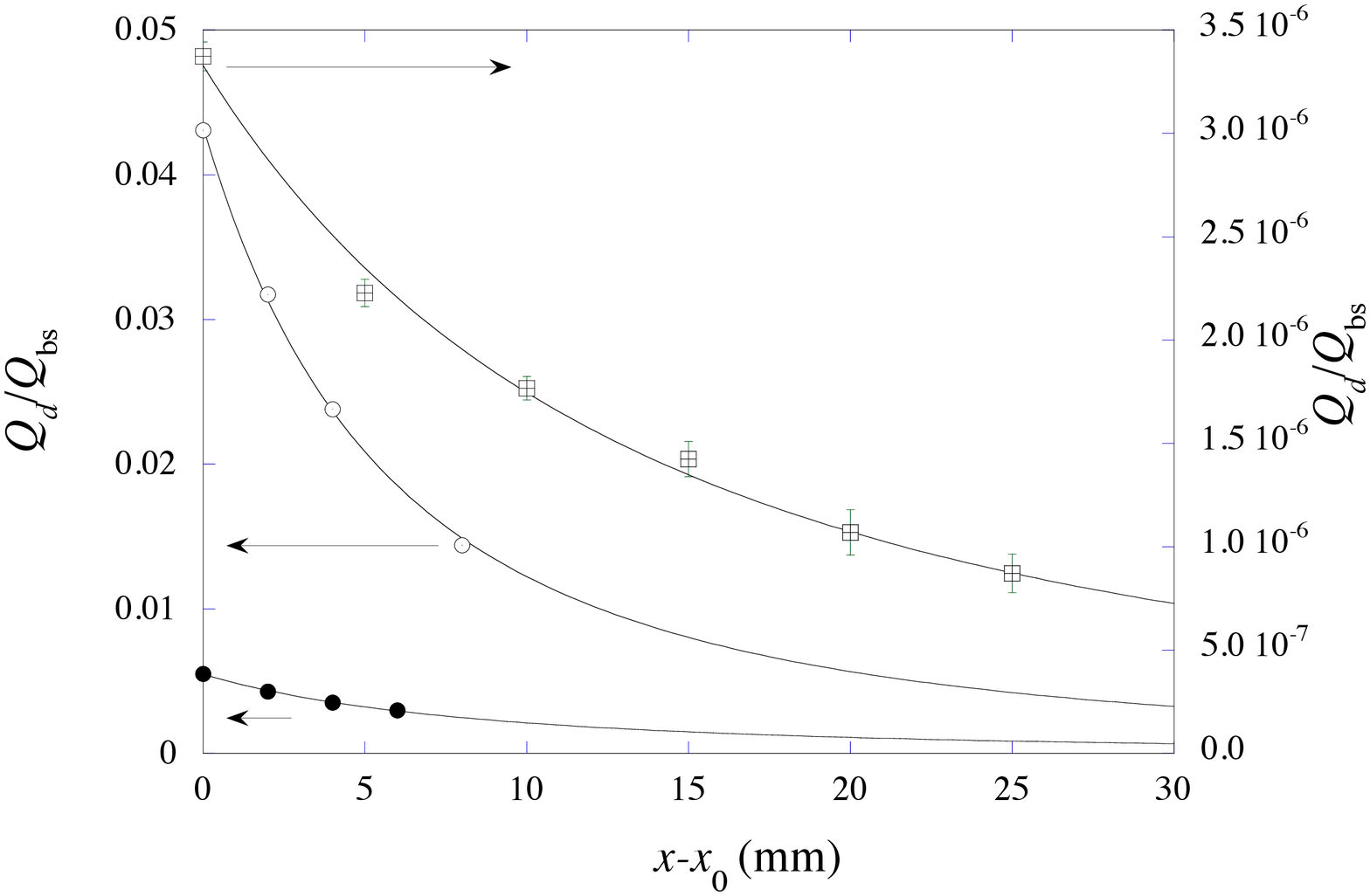}
\caption{\small Slope $Q_{d}/Q_\mathrm{bs}$ as a function of the detector position with respect to the Tm:YAG crystal for different spectral ranges and excitation methods. Circles (left scale): CL measured with the Si- (open symbols) and InAs detectors (closed symbols). Squares (right scale): RL measured with the Si detector.
\label{fig:distanze}}
\end{figure}
These quantities are also measured for different crystal-to-detector distances in order to determine the solid angle $S/d^2$. 

As an example, we show in \figref{fig:distanze} the ratio $Q_{d}/Q_\mathrm{bs}$ as a function of relative crystal-to-detector distance $x=d-x_{0}$ obtained for the Tm:YAG crystal excited with both electron-beam and X-rays. The absolute distance is defined within an arbitrary offset $x_{0}$ that depends on the physical detector mounting. Assuming that the crystal can be treated as a point source, the data must obey the equation
\begin{equation}
\frac{Q_{d}(x)}{Q_\mathrm{bs}}= \left(
\frac{Q_{d,0}}{Q_\mathrm{bs}}
d^{2}\right) \frac{1}{\left(x+x_{0}
\right)^{2}}= \frac{a}{\left(x+x_{0}
\right)^{2}}
\label{eq:d2}\end{equation}
The parameters $a=(Q_{d,0}/Q_\mathrm{bs})
d^{2}$ and $x_{0}$ are obtained by fitting \eqnref{eq:d2} to the experimental data.
The fit goodness confirms the validity of the point source approximation and allows us to determine the solid angle.

Finally, a more accurate determination of LY in the extended wavelength range 
can only be obtained by recording the emission spectrum  $I(\lambda)$ in the whole band.  Moreover, the spectral analysis is necessary to identify the levels responsible for the emission in the different bands and to ascertain how they are populated by the particle passage. By so doing, we can also compare our results with literature results in the bands, in which they are available, obtained with several different excitation techniques, including optical excitation~\cite{Gruber1989,Burdick1994,Gruber2004}.

Unfortunately, the complete spectrum in the extended wavelength band from DUV to mid IR cannot be obtained with a single spectrometer. As mentioned in Sect.~\ref{sect:app}, we used two spectrometer types covering different spectral ranges. The procedure to merge the suitably normalized spectra is described in the Appendix.

\begin{figure}[t!]
\centering
\includegraphics[width=\textwidth]{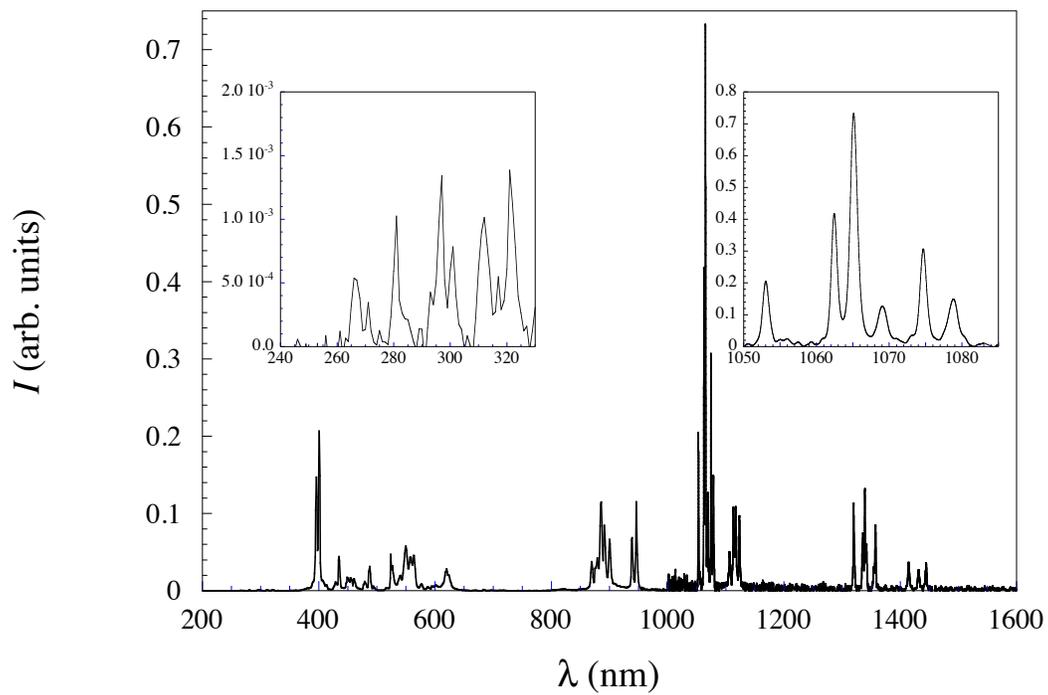}
\caption{\small Broadband CL of Nd:YAG. For $\lambda <1000\,$nm the spectrometer resolution is $\approx 2\,$nm. For $\lambda> 1000\,$nm the interferometer resolution is $4\,$cm$^{-1}.$\label{fig:spettroNd}}
\end{figure}

\begin{figure}[b!]
\centering
\includegraphics[width=\textwidth]{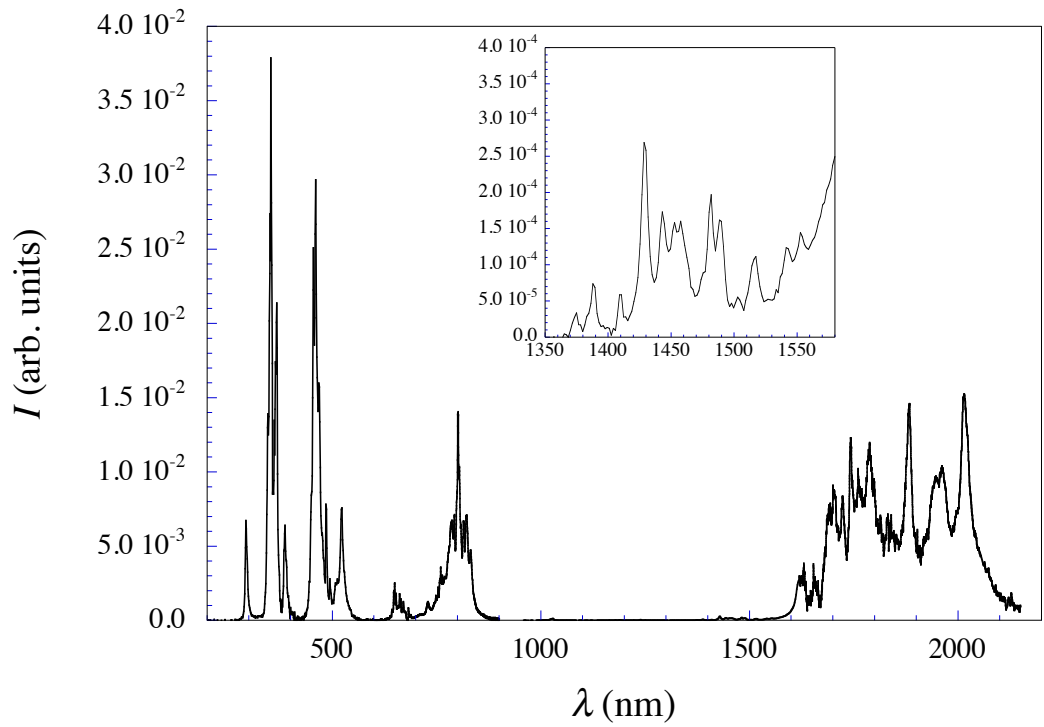}
\caption{\small Broadband CL of Tm:YAG. For $\lambda <1600\,$nm the spectrometer resolution is $\approx 2\,$nm. For $\lambda> 1600\,$nm the interferometer resolution is $4\,$cm$^{-1}.$\label{fig:spettroTm}}
\end{figure}

The CL spectra for Nd- and Tm- doped YAG crystals obtained in the extended spectral range are reported in \figref{fig:spettroNd} and \figref{fig:spettroTm}. In the $1.5-5\,\mu$m range for Nd:YAG and $2.2-5\,\mu$m range for Tm:YAG no lines have been observed. For this reason, these parts of the spectra are not displayed in the figures.

\clearpage
\subsection{Nd:YAG spectrum}

The visible portion of our extended CL spectrum for Nd:YAG compares favorably with the spectra in the same region obtained by several groups that excited the crystals with different techniques (laser~\cite{Venikouas1984};  CL~\cite{Gulyaeva2013}; ion beam~\cite{brooks2002,khanlary2005}; $\alpha$-particle~\cite{Seregina2013,yanagida2011}; synchrotron~\cite{Ning2007}).

\begin{figure}[t!]
\centering
\includegraphics[width=\textwidth]{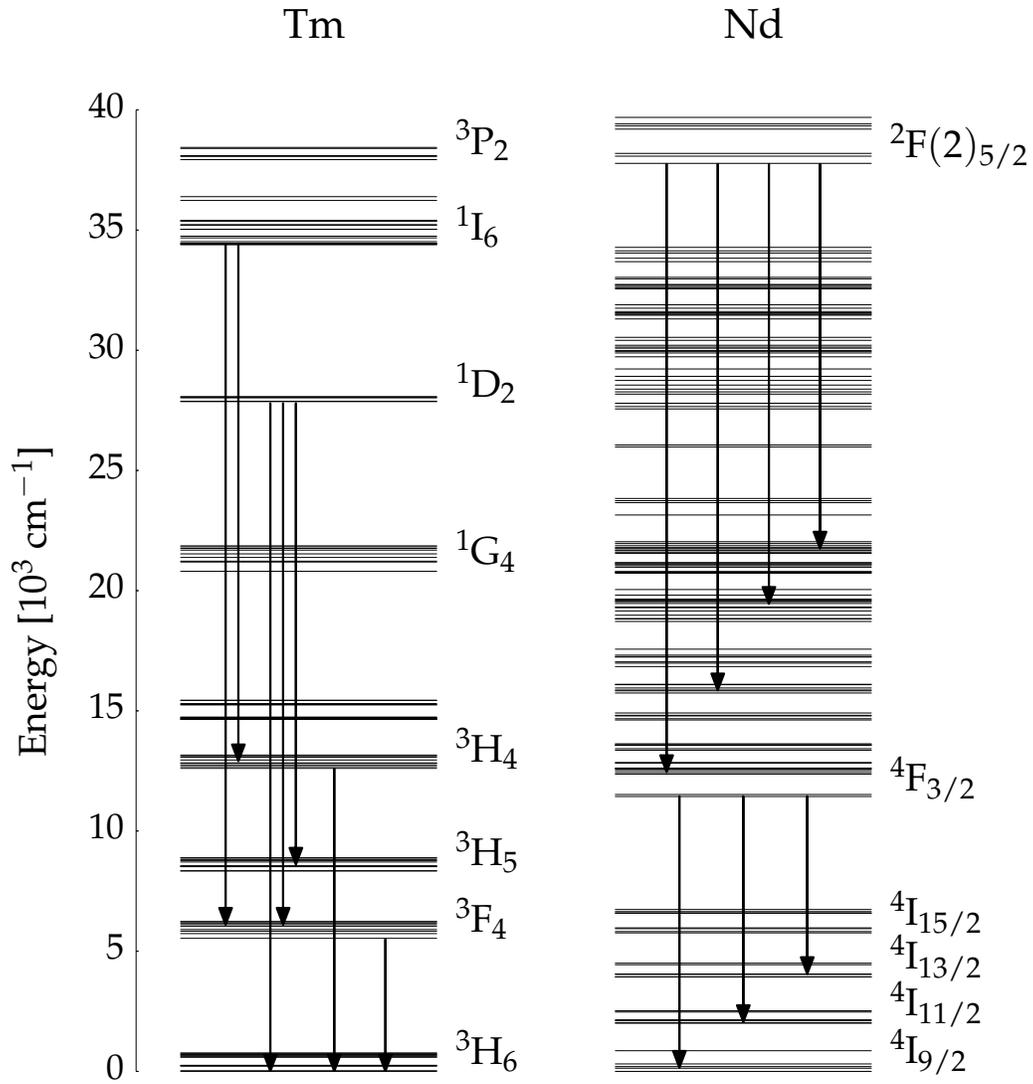}
\caption{\small Scheme of the Tm$^{3+}$~\cite{Gruber1989} and Nd$^{3+}$~\cite{Burdick1994} levels in YAG matrix. Only the strongest radiative transitions are shown.\label{fig:schemalivelli}}
\end{figure}

It is due to the transitions from the $4f^{3}$  manifold $^{2}\mathrm{F}(2)_{5/2}$  towards several lower lying manifolds that lie in the $11000-22000\,\mbox{cm}^{-1}$ wavenumber range and 
that are explicitly identified in literature~\cite{Ning2007}. The ultimate fate of all of these manifolds is to nonradiatively relax to the lowest of them, namely the $^{4}\mbox{F}_{3/2}$~\cite{Venikouas1984}. The energy level scheme of Nd is shown in \figref{fig:schemalivelli}.

The  ${}^{4}\mbox{F}_{3/2}$ manifold is responsible for all of the 27 lines we observed in the IR part of the spectrum. In agreement with literature~\cite{Singh1974,Burdick1994}, we have been able to identify all the Stark levels of the manifolds involved in the transitions. The ${}^{4}\mbox{F}_{3/2}\rightarrow {}^{4}\mbox{I}_{9/2}$,  ${}^{4}\mbox{F}_{3/2}\rightarrow {}^{4}\mbox{I}_{11/2}$, and $^{4}\mbox{F}_{3/2}\rightarrow {}^{4}\mbox{I}_{13/2}$  transitions lie in the $850-950\,$nm, $1040-1110\,$nm, and $1350-1450\,$nm ranges, respectively. The  $^{4}\mbox{F}_{3/2}\rightarrow {}^{4}\mbox{I}_{15/2}$ transitions lying in the  $1730-2100\,$nm range are very faint because of their little branching ratio and are obscured by the experimental noise. 
The emission from the ${}^{4}\mbox{I}_{J}$, as for the other manifolds, has not been observed mainly because it is quenched due to the large phonon energies of the oxide matrix,

The knowledge of the spectrum intensity $I(\lambda)$ allows us to compute the LY in a given wavelength band $\Delta \lambda_{i}$, according to the formula 
\begin{equation}
\mathrm{LY}_{i}\propto 
\int\limits_{\Delta\lambda_{i}} I(\lambda) \lambda\, \mathrm{d}\lambda
\label{eq:LY}\end{equation}
We found a ratio of the number of photons $N_\mathrm{IR} $ emitted in the IR band $\Delta\lambda_{\mathrm{IR}}$ $(800-1500\,\mbox{nm})$ to that of photons $N_{v}$ emitted in the visible range $\Delta \lambda_{v}$ $(390 -700\,\mbox{nm})$ 
 \[
 \alpha_\mathrm{Nd}=  \frac{N_\mathrm{IR}}{N_{v}}= 5.2
\]
with an accuracy better than 5\,\%.

There is a strong indication in literature that the $^{2}\mathrm{F}(2)_{5/2}$ manifold nearly completely relaxes by radiative decay because the multiphonon relaxation lifetime is more than two orders of magnitude longer than the radiative lifetime and because Nd-Nd energy transfer processes involving this manifold at the concentration of our experiment are negligible~\cite{Venikouas1984}, as confirmed by the fact that the lifetime difference in two single crystals of very different concentration is small~\cite{Seregina2013,yanagida2011}.

Moreover, according to the branching ratio values found in literature~\cite{Seregina2013}, $\approx 95\,\%$ of the radiative emission from the 
 $^{2}\mbox{F}(2)_{5/2}$ manifold lies in the visible range. Actually, we do not observe any IR emission attributable to this manifold and the UV emission originating from the ${}^{2}\mbox{F}(2)_{5/2}\rightarrow {}^{4}\mbox{I}_{J}$ transitions is a negligible fraction of the total one, as can be ascertained by inspecting the left inset of~\figref{fig:spettroNd}.
 Owing to the low dopant concentration and small size of our sample, we can estimate the visible reabsorption to be of a few \%, at most~\cite{Dong2005}.  As a consequence, the ratio $\alpha$ is a quite accurate estimate of the ratio of the numbers of photons emitted by the  $^{4}\mathrm{F}_{3/2}$ and ${}^{2}\mathrm{F}(2)_{5/2}$ manifolds, respectively.

In contrast with the $^{2}\mathrm{F}(2)_{5/2}$ manifold, the $^{4}\mathrm{F}_{3/2}$ manifold  shows a nonradiative loss channel due to the resonant energy transfer~\cite{Blatte1973} 
\[  {}^{4}\mathrm{F}_{3/2} + {}^{4}\mathrm{I}_{9/2} \rightarrow {}^{4}\mathrm{I}_{15/2} + {}^{4}\mathrm{I}_{15/2}
\]
At the concentration of our sample, this kind of loss amounts to $\approx 10\,\% $~\cite{Dong2005}.
Owing to these estimates, we can draw the conclusion that $\alpha_\mathrm{Nd}=5.2$ is also a direct measure of the ratio of the numbers of ions excited in these two manifolds. However, if  electrons had the effect to only excite the upper 
$^{2}\mathrm{F}(2)_{5/2}$ manifold, the previous arguments would lead us to the conclusion that we  should get $\alpha_\mathrm{Nd}\approx 1,$ as demonstrated by Venikouas {\em et al.} who directly excited the $^{2}\mathrm{F}(2)_{5/2}$ manifold in a sample similar to our with a quadrupled Nd:YAG laser pumping at $\lambda = 266\,$nm~\cite{Venikouas1984}. 
Therefore, we are forced to draw the conclusion that, by exciting the crystals with particles, the $^{4}\mbox{F}_{3/2}$ is populated by the $^{2}\mbox{F}(2)_{5/2}$ direct relaxation by only $\approx 20\,\%$. Thus other excitation mechanisms have to be active.

Whereas the processes leading to the population of the charged particle-excited $^{2}\mbox{F}(2)_{5/2}$ as a consequence of the relaxation of the high-lying $4f^{2}5d$ levels are known~\cite{Ning2007,lempicki1995} (i.e, multiphonon cascading through the lowest $4f^{2}5d$ level,  then to the $^{2}\mbox{F}(2)_{7/2}$, and, finally, to the $^{2}\mbox{F}(2)_{5/2}$), the processes of populating the IR emitting $^{4}\mbox{F}_{3/2}$ manifold, at the best of our knowledge,  are not studied at all.
We can, tentatively, suggest that, among the processes leading to particle excitation of the $^{4}\mbox{F}_{3/2}$ manifold, the most important could be (i) direct electromagnetic excitation from the RE ion ground state, (ii)  excitation by secondary electrons, and (iii) particle induced lattice distortions that relax by phonon emission.

\subsection{Tm:YAG spectrum}
The extended CL spectrum of the Tm:YAG crystal is reported in \figref{fig:spettroTm}.  
Also in this case, the visible and very near IR region of our spectrum (up to $\lambda \approx 850\,$nm) favorably compares with those obtained by Yanagida {\em et al.} by exciting Tm:YAG~\cite{Fujimoto2013}, and Tm:LuAG- and Tm:YAP~\cite{yanagida2013} crystals with $\gamma$-rays. Tm shows a scheme of the $4f$ energy levels that is far less rich than Nd. 
Most of its manifolds are well separated in energy. As their multiphonon relaxation rate is low, they mainly decay radiatively thereby making more difficult the identification of the transitions from the lines observed in the spectrum.
As different levels may emit at the same wavelength or at very near wavelengths but have different lifetimes, we have been able  to resolve the identification ambiguity by analyzing the time evolution of the PD signal in the desired wavelength band. The identification of the manifolds involved in the several transitions and the branching ratio we have obtained agree well with the results obtained by the groups that selectively excited the same manifolds by using  lasers~\cite{Fei2013,Thomas2013}. 
The manifolds responsible for the visible emission are: $^{1}\mbox{I}_{6},$ $^{1}\mbox{D}_{2},$ and $^{1}\mbox{G}_{4}$.
The energy level scheme for Tm is shown in \figref{fig:schemalivelli}.

Our CL spectrum for $\lambda >1600\,$nm is identical with that obtained with laser excitation~\cite{Thomas2013}, well reproduced by the theoretical computation due to Fei {\em et al.}~\cite{Fei2013}, and is associated with the transition from the  $^{3}\mbox{F}_{4}$ manifold  towards the $^{3}\mbox{H}_{6}$ ground state. We have estimated the lifetime of the emitting manifold to be $\approx 6.6\,$ms. This result is in reasonable agreement with literature data, whose spread (from $\approx 4\,$ms to $\approx 12\,$ms) is, unfortunately, very large~\cite{antipenko1978,Thomas2013}.

The emission in the $1350-1550\,$nm range, shown in the inset in \figref{fig:spettroTm}, and that in the $800-850\,$nm range are ascribed to the transitions ${}^{3}\mbox{H}_{4}\rightarrow {}^{3}\mbox{F}_{4}$ and ${}^{3}\mbox{H}_{4}\rightarrow {}^{3}\mbox{H}_{6},$ respectively.
As these emissions are originating from the same manifold, they share the same time evolution, from which we determined a lifetime of  $\approx 110\,\mu$s, in agreement with literature results obtained in a sample with similar dopant concentration~\cite{barnes2002}.

We can now give an estimate of $\alpha .$ In this case, we have chosen  $\Delta \lambda_\mathrm{IR}$ to span the $1600-2200\,$nm range in order to account for the emission of the first excited $^{3}\mbox{F}_{4}$ manifold. For $\Delta\lambda_{v}$ we have chosen the $280-850\,$nm range that accounts for the emission of all other higher lying manifolds.
We get
\[
\alpha_\mathrm{Tm} \approx 6.4
\]
Contrary to the Nd:YAG case, the value of the $\alpha_\mathrm{Tm} $ parameter is not related  in a simple way to the number of ions excited by the charged particles in specific manifolds because their quantum efficiency is hardly known. Actually, 
the Tm energy level scheme is such that to favor concentration-dependent energy transfer processes that are in competition with the channel of radiative decay. The presence of many and important energy transfer processes (namely, cross relaxation (CR)) has been confirmed both by computation and observation for Tm in YLF matrix~\cite{tkachuk2001}.
We believe that the same energy transfer processes also occur in Tm:YAG because of the following experimental observations. First of all, we measured  a lifetime value of $\approx 110\,\mu$s for the $^{3}\mbox{H}_{4}$ manifold that is much shorter than the value $\gtrsim 500 \,\mu$s observed and computed in low dopant concentration Tm:YAG crystals~\cite{barnes2002,Thomas2013}. An efficient cross relaxation process affects the $^{3}\mbox{H}_{4} $ manifold according to the scheme
\[
^{3}\mbox{H}_{4} + {}^{3}\mbox{H}_{6}\rightarrow {}^{3}\mbox{F}_{4} + {}^{3}\mbox{F}_{4}
\]
that leads to a nonradiative increase of the population of the IR emitting ${}^{3}\mbox{F}_{4}$ manifold.

Secondly, the presence of energy transfer processes is confirmed by comparing the visible CL spectrum with the visible spectrum obtained by exciting the crystal with a quadrupled Nd:YAG laser at $266\,$nm. The two spectra are practically identical because they are both originated by the relaxation of the $^{1}\mbox{I}_{6}$ manifold. The laser directly populates the $^{3}\mathrm{P}_{J}$ manifolds whereas charged particles populate the same manifolds by nonradiative relaxation of the $4f^{11}5d$ levels. Then, the $^{3}\mbox{P}_{J}$ nonradiatively relax to the $^{1}\mbox{I}_{6}.$ As we observe visible emission originating from the $^{1}\mbox{D}_{2}$ and $^{1}\mbox{G}_{4} $ manifolds, we have to conclude that the latter two manifolds are populated by cross relaxation because  no radiative transitions from the $^{1}\mbox{I}_{6} $ manifold towards  them are observed and because multiphonon relaxation is negligible owing to the large energy gap between the manifolds.
In third place, the great influence of CR processes in our sample 
is confirmed by the fact that the emission from ${}^{1}\mbox{D}_{2}$ and ${}^{1}\mbox{G}_{4}$ is much more intense than in a sample of much lower concentration~\cite{Fujimoto2013}. 
In any case, as the quantum efficiency of the $^{3}\mbox{F}_{4}$ manifold is $\approx 100\,\%$~\cite{Fei2013}, the number of photons emitted in the $\Delta \lambda_\mathrm{IR}$ band equals the number of ions populating this manifold via the several aforementioned mechanisms.

We can conclude this section on the Tm:YAG spectrum in the extended wavelength range by remarking that energy transfer processes are a very important channel to populate the low energy, IR emitting manifolds. Moreover, their influence does not allow us to precisely identify and quantify the remaining channels.

As a final remark, we note that the characteristic emission~\cite{anedda2006,rydberg2013} due to the Fe$^{3+}$ contamination of the YAG matrix we obtained by exciting the crystal with a quadrupled Nd:YAG laser at $266\,$nm is absent under electron-beam excitation.

\subsection{Absolute calibration of LY with X-rays}

As a result of the analysis of the Nd:YAG and Tm:YAG luminescence we have demonstrated that the low lying, IR emitting manifolds are very efficiently populated, roughly a factor 5 more than the higher lying levels that emit in the visible range. The $\alpha$ parameter defined above represents the IR LY relative to the visible one.  However, the knowledge of the absolute value of the light yield is required to design a detector. Therefore, we have compared the luminescence of our samples with that of a reference crystal whose light yield is known.

Unfortunately, absolute LY measurements are usually carried out with X- or $\gamma$-rays excitation. As a consequence, we also measured the RL of our samples by exciting them with X-rays of energy of a few tens of keV, produced by converting the energy of the electrons impinging on the Ta film.

For the calibration purposes, we have measured the RL of a Pr:LuYaG  crystal whose LY is reported to be $2.7\times 10^{4}\,$ph/MeV in the $300-450\,$nm range when excited with $662\,$keV $\gamma$-rays~\cite{drozdowski2014}. From the our RL spectrum, by taking into account the small non-proportionality of the LY over a wide energy range ~\cite{swiderski2009light}, we obtain $\mathrm{LY}=3.3\times 10^{4}\,$ph/MeV in the optical range of the Si detector we used~\cite{chiossi2017}.

\begin{figure}[b!]
\centering
\includegraphics[width=\textwidth]{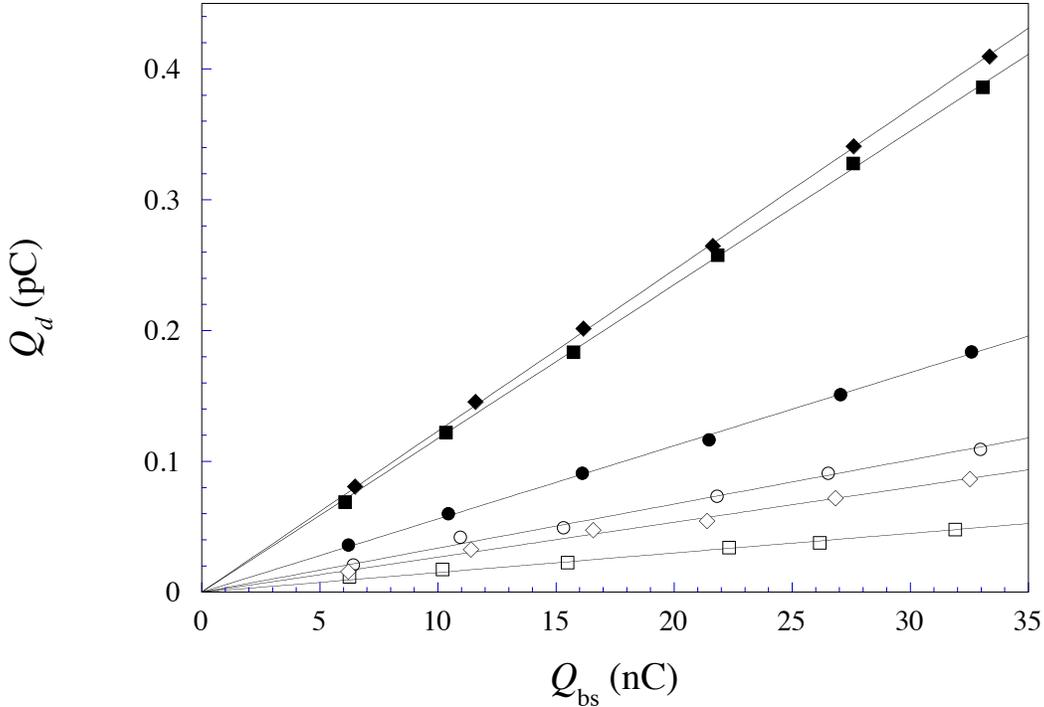}
\caption{\small $Q_{d}$ vs $Q_\mathrm{bs} $ for the X-rays excited Tm:YAG (circles), Nd:YAG (diamonds), and Pr:LuYAG (squares) crystals.
Closed symbols: response of the Si detector irradiated with both luminescence light and X-rays. Open symbols: contribution due to X-rays only. 
\label{fig:NdTmPrLinearitaX}}\end{figure}

Also in the case of X-rays we have a direct proportionality between $Q_{d}$ and the X-rays intensity that is proportional itself to $Q_{\mathrm {bs}}$, as shown in \figref{fig:NdTmPrLinearitaX}.

The calibration consists in determining the $k$ constant in \eqnref{eqn:LY5}. As the detector quantum efficiency $\eta$ is practically constant over the wavelength range of Pr emission, we obtain $k$ from a measurement of the parameter $a$ and from the LY as
 \begin{equation}
k= \left(\frac{a}{\mbox{LY}}\right)\frac{1}{4\pi S\eta}\label{eq:k}\end{equation}

\noindent The LY for the Nd:YAG and Tm:YAG crystals in the extended range can now be obtained from measurements of $a=(Q_{d,0}/Q_\mathrm{bs})d^{2}$ and from the emission spectra. Unfortunately, the intensity of RL is much weaker than that of CL and RL spectra can only be recorded with CCD spectrometers.

The RL and CL spectra we obtained for both Nd:YAG and Tm:YAG are identical except for the different resolution of the different spectrometers used for the IR range. In particular, for the Nd:YAG case, in the $200-1000\,$nm range we observe that the relative intensities of the  emissions stemming from the ${}^{2}\mbox{F}(2)_{5/2}$ and from the ${}^{4}\mbox{F}_{3/2}$ are are the same in both CL and RL spectra. As the spectrum for $\lambda>1000\,$nm is only originated by the same ${}^{4}\mbox{F}_{3/2}$ manifold responsible for the emission around $ \lambda\approx 800\,$nm and the branching ratios do not depend on the excitation type, we can use the more accurate CL spectra in order to compute the LY. This conclusion is supported by the argument that ionizing radiation (X-rays or electrons) may originate similar cascading processes of energy degradation~\cite{lempicki1995}.

In the case of Tm:YAG the emission from ${}^{3}\mbox{F}_{4}$ falls beyond the reach of the CCD spectrometers. Thus, only the CL spectrum can be recorded with the FT-IR interferometer because the RL is too weak. In any case, owing to the Nd:YAG results and the relative arguments, we safely assume that CL spectra can be used to compute the LY.  

The results are reported in \tabref{tab:LYNdTm}.
\begin{table}[t!]
\centering
\caption{LY for Nd:YAG and Tm:YAG crystals.}
\begin{tabular}{c|c|c}
\toprule
\multicolumn{3}{c}{Nd:YAG 1.1\% at.}\\
\midrule
Manifold & Optical range [nm] & LY [$10^{3}\,$ph/MeV] \\
\midrule
$^2\mathrm{F}(2)_{5/2}$ &  390$-$650      &       9.5  \\
$^4\mathrm{F}_{3/2}$&  800$-$1500 &   50   \\
\toprule
 \multicolumn{3}{c}{Tm:YAG 4.4\% at.}\\
\midrule
Manifolds & Optical range [nm] & LY [$10^{3}\,$ph/MeV]\\
\midrule
$\mathrm{^1I_6,{}^1D_2,{}^1G_4,{}^3H_4}$ &  280$-$900     &       7.0 \\
$\mathrm{^3F_4}$ &  1600$-$2300   &       45    \\
\bottomrule
\end{tabular}
\label{tab:LYNdTm}
\end{table}
Their accuracy is estimated to be of the order of $15\,\% .$
For the Nd:YAG crystal we obtain $\mathrm{LY}\approx 10^{4}\,$ph/MeV in the visible range 
that compares very favorably with literature results. Actually, Yanagida {\em et al.} have reported a UV-visible $\mathrm{LY}=(11.0 \pm 1.1)\times 10^{3}\,$ph/MeV for 662\,keV $\gamma$-ray both for Nd:YAG 1.1\% at.~\cite{yanagida2011} and for Tm:YAG 0.5\% at. crystals~\cite{Fujimoto2013}. 
The small discrepancy between our result for Nd:YAG and that of Yanagida {\em et al.} is probably due to the difference in the energy 
of the ionizing radiation.

For Tm:YAG we obtain a significantly smaller value than Yanagida {\em et al.}. The reason can mainly be attributed to the much larger dopant concentration of our sample for which the cross-relaxation rate is expected to be much larger. 

On the other hand, the manifolds of lower energy emit a much larger photon number, $5\times 10^{4}\,$ph/MeV in the Nd:YAG case and $4.5\times 10^{4}\,$ph/MeV in the Tm:YAG case. 
These large numbers of ph/MeV emitted by the crystals in the IR range might attributed, as previously suggested, to the high probability of populating the lower energy levels of the dopant by means of several physical processes including cross-relaxation.

\section{Conclusions}\label{sect:conc} 
In this work we 
report the CL and RL of Nd- and Tm:YAG crystals in a wavelength range particularly extended toward the near or mid IR. The motivation of our study is the possibility to develop a low-threshold, low-rate particle detector based on the Infrared Quantum Counter scheme. 

The spectral analysis of CL and RL suggests that the low-lying levels emitting in the IR band are directly populated from the ground state by the ionizing radiation. 

We estimate that the IR light yield of the investigated crystals is a factor $\sim 5$ larger than in the visible range. In particular, we get an IR light yield $\approx 5\times 10^{4}\,$photons/MeV for both crystals stemming from the metastable manifolds we are interested in for the IRQC scheme. This piece of information is very useful for the detector design. 

We believe that the present work contributes a further step in the understanding of the particle excitation of lower lying levels in active materials. The goal of our future investigations is to identify the best combination of crystal and dopant species to achieve the highest possible light yield in the IR region taking into account the best upconversion schemes.

\section*{Acknowledgments}
The research is sponsored by Istituto Nazionale di Fisica Nucleare (INFN) within the AXIOMA project.
The authors gratefully acknowledge useful discussions with Prof. M. Tonelli of the University of Pisa and the technical assistance of Mr. L. Barcellan and E. Berto.

\clearpage

\appendix
\setcounter{figure}{0}

\section{Merging spectra obtained in two hardly overlapping bands}
In our setup we cover the wavelength band from $200\,$nm to $1000\,$nm with CCD spectrometers and the band from $900\,$nm up to $5\,\mu$m with the FT-IR interferometer endowed with either a InGaA, or a InAs, or a InSb photodetector. In order to obtain a spectrum over the extended wavelength range we have devised a procedure to merge the spectra recorded in the two different bands. 

The main requirement for this procedure to be valid is that the spectra intensity over all wavelengths is directly proportional to the amount of energy released in the crystals, as we have experimentally verified (see, for instance, \figref{fig:LinTmegun} and \figref{fig:NdTmPrLinearitaX}). Thus, the spectrum shape is independent of $Q_\mathrm{bs}$. 

The procedure is based on the relative measurement of the light yields in the two distinct wavelength bands, which we conventionally term $\Delta \lambda_{1}$ and $\Delta \lambda_{2}$. The normalization factor of the two spectra is
\begin{equation}
\mathcal{F}= \frac{\int\limits_{\Delta \lambda_{1}} 
I_{1}(\lambda)\lambda\,\mathrm{d}\lambda
}{\int\limits_{\Delta \lambda_{2}} 
I_{2}(\lambda)\lambda\,\mathrm{d}\lambda
}\,\left(\frac{\mbox{LY}_{2}}{\mbox{LY}_{1}}\right)
\label{eq:F}\end{equation} 
in which $I_{1}$ and $I_{2}$ are the unnormalized spectra recorded in the respective bands. The light yields are computed with the help of \eqnref{eqn:LY5} from the measured values of $a=(Q_{d,0}/Q_\mathrm{bs})d^{2}$ and from the integrated spectra.

The determination of the LY's is easy if the product $T(\lambda)\eta(\lambda)$ of the PD and filter combination with which $a$ is measured is different from zero in a wavelength band completely contained within the working band of the spectrometer used.
In this case, only the wavelengths belonging to the recorded spectrum $I(\lambda)$ contribute to the measured $Q_{d}$ value.
Moreover, in this way we get rid of the necessity to measure the conversion factor $k$ in \eqnref{eqn:LY5}.

A problem, however, arises because the optical range of Si detector used to measure $Q_{d}$ in the shorter wavelength region is a little broader than than that of the CCD spectrometer range. Whereas the latter extends up to $1000\,$nm, the sensitivity of the former is non negligible up $\approx 1100\,$nm, at least. This issue only affects the normalization procedure for Nd:YAG as it shows significant emission in this restricted wavelength range. On the contrary, the procedure for Tm:YAG is not affected because it negligibly emits in the same region.

The simplest way to overcome this problem would be to use suitable filters of known transmittance $T(\lambda)$. 
However, we have devised a more general procedure in case the right filters are not available. It is based on the realization that the LY in the overlapping region of the optical ranges of the two PD's must be the same, independently of the PD used.
We obtain the $ \mbox{LY}_{c} $ in the $1000-1100\,$nm 
range as a fraction of the total, uncalibrated infrared LY. 
As the $\mbox{LY}_{c}$ has to be independent on the detector type and the spectrum is known, we can compute the response $a_{c}$
of the Si detector in the region of interest. Now, its response $\tilde a$ in the spectrometer working range is simply obtained as $\tilde a = a_{m} -a_{c},$ where $a_{m }$ is the measured Si detector response over its entire range. Once $\tilde a $ is known, \eqnref{eqn:LY5} allows us to compute the LY in the range of the CCD spectrometer, which is finally used in \eqnref{eq:F} to obtain the normalization factor $\mathcal{F}$.

It turns out that the following equation must hold true
\begin{equation}
\frac{Q_{d}}{Q_\mathrm{bs}}d^{2}\propto \int I(\lambda)T(\lambda)\eta(\lambda) \lambda\,\mathrm{d}\lambda 
\label{eq:validazione}\end{equation}
 
\begin{figure}[t!]
\centering
\includegraphics[width=\textwidth]{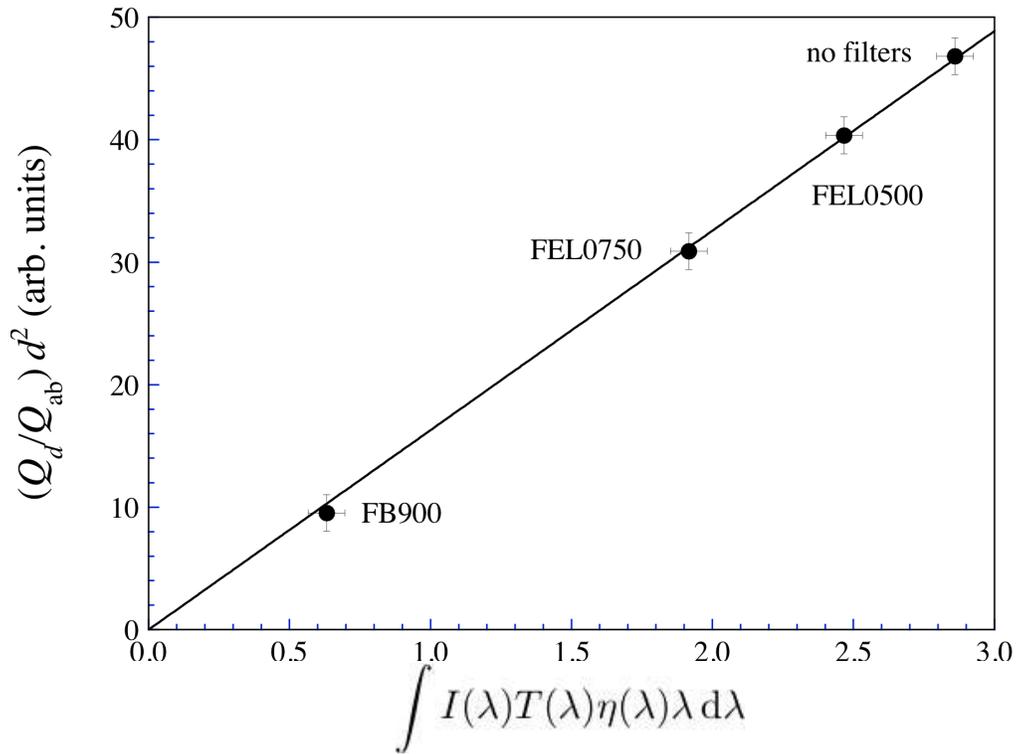}
\caption{\small Si detector response in different wavelength bands vs numerical integration of the weighted normalized spectrum over the same bands. The bands are defined by the filters used. FEL: longpass filters with cut-in wavelength  500$\,$nm and $750\,$nm, respectively. FB: passband filter centered at $\lambda=900\,$nm with width $40\,$nm.
\label{fig:validazione}}
\end{figure}

\noindent In \figref{fig:validazione} we show the detectors' response $(Q_{d,0}/Q_\mathrm{bs})d^{2}$ over several wavelength bands defined by suitable filters vs the numerical integration of the normalized spectrum over the same bands.
The direct proportionality between the two different determinations of the same quantity is the validation of the normalization procedure.

\end{document}